\documentclass[11pt]{revtex4}

\usepackage{hyperref}

\usepackage{hyperref}
\usepackage{graphicx}
\usepackage{dcolumn}
\usepackage{color}
\usepackage{bm}
\hypersetup{dvips,dvipdfm,colorlinks=true,urlcolor=magenta,filecolor=magenta,linktoc=page,citecolor=red,linkcolor=blue,bookmarks=true}
\usepackage{amsmath}
\usepackage{amsfonts}
\usepackage{amssymb}
\usepackage{epstopdf}
\usepackage{float}

\input epsf

\begin{document}
	
	\title{Universe consisting of diffusive Dark fluids: Thermodynamics and stability analysis}

	\author{Subhayan Maity\footnote {maitysubhayan@gmail.com}}
	\affiliation{Department of Mathematics, Jadavpur University, Kolkata-700032, West Bengal, India.}
	\author{Pritikana Bhandari\footnote {pritikanab@gmail.com}}
	\affiliation{Department of Mathematics, Jadavpur University, Kolkata-700032, West Bengal, India.}
	
	\author{Subenoy Chakraborty\footnote {schakraborty.math@gmail.com}}
	\affiliation{Department of Mathematics, Jadavpur University, Kolkata-700032, West Bengal, India.}


	\maketitle

	
	
	
	\section{Abstract}
	The present work deals with homogeneous and isotropic FLRW model of the Universe having a system of non-interacting diffusive  cosmic fluids with barotropic equation of state (constant or variable equation of state parameter). Due to diffusive nature of the cosmic fluids, the divergence of the energy momentum tensor is chosen to be proportional to the diffusive current. The thermodynamic stability analysis of individual fluids is done and the stability conditions are expressed as restrictions on the equation of state parameter. \\
	 
	{\bf Keywords\,:} Thermodynamical equilibrium, Diffusive cosmic fluid, Equation of State, Thermodynamic Properties.\\\\
	PACS Numbers\,: 95.30.Sf, 98.80.Cq, 66.10.Cb, 74.25.Bt\\\\ 
	
	\section{Introduction} 
	The greatest challenge of standard cosmology today is to accommodate the recent observational predictions \cite{Perlmutter:1998np,Riess:1998cb,deBernardis:2000sbo,Percival:2001hw,Spergel:2003cb,Jain:2003tba,Tegmark:2003ud,Eisenstein:2005su,Komatsu:2010fb}. The modern cosmology is facing the challenging issue of explaining the present accelerated expansion of the universe. In the framework of Einstein gravity, cosmologists are speculating some hypothetical exotic matter (known as dark energy (DE) having large negative pressure) to explain this accelerating phase. It is estimated \cite{Ade:2015xua} that about 70 percent of the cosmic fluid consists of this unknown DE component. The simplest as well as the common candidate for this dark fluid is the cosmological constant  (the zero point energy of the quantum fields). Although a large number of available observational data are in support of this cosmological parameter as a DE candidate but there are severe drawback of it  at the interface of Cosmology and Particle Physics : the cosmological constant problem \cite{Weinberg:2000yb,Padmanabhan:2002ji} and the coincidence problem \cite{Steinhardt:2003st}. As a result there are several alternative proposed dynamical DE models  into the picture. These DE models have been studied for the last several years \cite{Amendola1}, yet the cosmological constant is still the best observationally supported DE candidate. However, these different dynamic DE models cannot be compared from observational view point as these models try to adjust the data seamlessly. As a result, cosmologists have been trying with interacting DE model to have a better understanding of the mechanism of this cosmic acceleration.
	\par From recent past, interacting dark fluid models have been receiving much attention as they can provide small value of the cosmological constant and due to  their ability of explaining the cosmic coincidence problem \cite{Bolotin:2013jpa,Wang:2016lxa,Sharov:2017iue}. Moreover, recent observed data \cite{Salvatelli:2014zta,Sola:2015wwa,Sola:2016ecz,Nunes:2016dlj,Kumar:2016zpg,vandeBruck:2016hpz} favor interacting dark fluid models and it is possible to have an estimate of the coupling parameter in the interaction term by various observations\cite{Nunes:2016dlj,Kumar:2016zpg,vandeBruck:2016hpz,Cai:2015zoa,Yang:2014gza,Yang:2014hea,Yang:2016evp,Xia:2016vnp,Caprini:2016qxs,Murgia:2016ccp}. Further,the cosmologists are of the opinion that these interacting dark fluid models may have the solution to the current tensions on $\sigma_8$ and the local value of the Hubble constant $(H_0)$ \cite{Kumar:2016zpg,vandeBruck:2016hpz,Cai:2015zoa,Yang:2014gza,Yang:2014hea,Yang:2016evp,Xia:2016vnp,Caprini:2016qxs,Murgia:2016ccp,Pourtsidou:2016ico}. Moreover, it is speculated that cosmological perturbation analysis may be affected by the interaction terms and consequently, the lowest multipoles of the CMB spectrum \cite{Zimdahl:2005bk,Wang:2006qw} should have an imprint of it.
	\par On the other hand, the unknown nature of DE may have some clue from the thermodynamic laws which  are applicable to all types of macroscopic systems and are based on experimental evidence. However, unlike classical mechanics or electromagnetism, thermodynamical analysis can not predict any definite value for observables, it may only give limit on physical processes. So it is reasonable to believe that thermodynamical study of dark cosmic fluid may indicate some unknown character of it. Investigation in this direction has been initiated recently by Barboza et al \cite{Barboza:2015rsa}. But their stability conditions demand that the DE should have constant (-ve) equation of state and it is not supported by observation.
	\par Subsequently, Chakraborty and collaborators \cite{Bhandari:2017cow,Bhandari:2018fon,Mamon:2018flx,Haldar:2017tqd} in a series of works have shown the stability criteria for different type of dark fluids and have presented the stability conditions (in tabular form) for different ranges of the equation of state parameter. The present work is an extension of the interacting dark fluid model. Here a particular realization of the non-conservation of the energy momentum tensor is used  with a diffusion of dark matter in a fluid of dark energy. The change of the energy-density is  proportional to the particle density. The resulting non-equilibrium thermodynamics is studied and stability conditions are explicitly determined. The paper is organized as follows: 
	Section  \ref{sec2}  deals with thermodynamical analysis of non-interacting cosmic fluids with constant equation of state . The stability criteria for non-interacting cosmic fluids with variable equation of state has been organized in Section  \ref{sec3} . Finally, the paper ends with a brief discussion and the conclusion from the result in dark energy range in Section  \ref{sec4} .
	
	\section{Thermodynamic analysis of non-interacting diffusive cosmic fluids having constant equation of state parameter} \label{sec2}
	
	The universe is assumed to have different types of non-interacting cosmic fluids (including dark energy and dark matter). Here it will be examined whether all kinds of diffusive fluids are thermodynamically stable or not  in an adiabatic universe . Let the equation of state of these cosmic fluids are barotropic in nature having explicit form (for i-th fluid )
	\begin{equation} 
	p_i = w_i\rho_i ,\label{1}
	\end{equation}
	where  $\omega_i$ ,a constant ,is the barotropic index of the fluid and  $p_i$ and $\rho_i$ are the pressure and  energy density of the fluid respectively. So the Friedmann equations for the  whole system  take the form 
	\begin{eqnarray} 
	3H^2 &=\sum_i\rho_i           \label{2}\\
	2\dot H+3H^2 &= - \sum_ip_i     \label{3}
	\end{eqnarray} 
	where $ H=a^{-1}\dot{a}$ is the  Hubble parameter ,    $a(t)$ is called the scale factor of the universe as all physical distance is scaled with same factor `` $a$ " due to homogeneity and isotropy of space-time. Now due to the diffusive nature of the fluids,~they do not obey the matter conservation equation (${T_i^{\mu \nu}}_{;\mu}= 0$), rather they follow, 
	\begin{equation}
	T_ i^{\mu \nu}{;\mu} = 3k^2N_i^\nu ,   \label{4}
	\end{equation}
	where  $N_i^{\nu}$ is the  current of diffusion corresponding to that fluid  and $ k$ is a constant. But for the whole universe,the total  stress-energy tensor is conserved i.e.(  $\sum_i{{T_i^{\mu \nu}}_{;\mu}= 0}$  )
	\begin{equation}
	\sum_iN_i^{\nu} = 0              \label{5}.            
	\end{equation}
	In simplest form, the conservation equation(non conservation) takes the form for FLRW universe  \cite{Amendola:2006dg,Bohmer:2009zz,Lima:2014hda} as, 
	\begin{equation}
	\frac{\partial \rho}{\partial t} + 3H(1+\omega_i)\rho_i = \gamma_ia^{-3}         \label{6}
	\end{equation}
	where, $\gamma_i$ is a constant for a particular fluid but it is different for different fluids with $\sum_i \gamma_i = 0$.
	Now, in the context of classical thermodynamics,the contribution to the energy of the  universe by i-th fluid is given by 
	\begin{equation}
	E_i=\rho_iv                      \label{7}
	\end{equation}
	where, $v = v_0a^3(t)$ is the physical volume of universe at given time $t$ and at present time $t_0$,the physical volume is $v_0$ with $a(t_0)=1$. According to the 1st law of thermodynamics(which is an energy conservation equation),
	\begin{equation}
	Td S_i=d E_i + p_i d v                     \label{8}.
	\end{equation}
	Integrating equation \,(\ref{6}) one  can easily find the expression for  energy density of  the i-th fluid as,
	\begin{equation}
	{\rho}_i={a^{-3(1+{\omega _i})}}\left [{\rho}_{i 0} + \gamma_i \int_{t_0}^{t}a^{3{\omega}_i} d t\right ]                              \label{9}
	\end{equation}
	where $\rho _{i 0}$ is the value of $\rho _i$ at $t=t_0$ i.e. at  the present time .The  energy density of i-th fluid, from  equation \,(\ref{9})   can be arranged as
	\begin{equation}
	d(\ln {\rho} _i )=-(1 + \omega _i)d(\ln v)+d[\ln (
	\rho _{i 0}+\gamma _i A_i)]              \label{10}
	\end{equation}
	where $A_i(t)=\frac{1}{v_0^{\omega_i}}\int_{t_0}^{t}v^{\omega_i}dt$.
	Now, taking entropy $S_i$ as a function of two independent thermodynamic  variables namely  volume($v$) and temperature ($T$) and also considering  $d S_i$ to be an exact differential, one obtains from equation \,(\ref{8}),
	\begin{equation}
	d \ln T = - \omega _i d \ln v+ \frac{w_i}{1+ \omega _i}d[\ln (\rho _{i 0}+\gamma _i A_i)]               \label{11}.
	\end{equation}  
	Now combining equations \,(\ref{10}) and  \,(\ref{11}),  a new relation among  energy density ,temperature and time can be written as  
	\begin{equation}
	\frac{T}{\rho _i v}=\beta{[\rho _{i 0}+\gamma _i A _i]}^{-\frac{1}{1+\omega _i}} ,         \label{12}
	\end{equation}
	where $\beta$ is an integration constant. Effectively at present time $t=t_0$,
	\begin{equation}
	\frac{T_0}{\rho _{i 0} v_0}=\beta \rho _{i 0}^{-\frac{1}{1+\omega _i}}      \label{13}.
	\end{equation}
	Now equations \,(\ref{12}) and  \,(\ref{13}) imply the evolution of  energy of the i-th fluid with time and temperature as,
	\begin{equation}
	E_i=E_{i 0}\left (\frac{T}{T_0}\right )\left [1+\frac{\gamma_i}{\rho _{i 0}}A_i \right ]^\frac{1}{1+\omega _i}              \label{14}.
	\end{equation}
	This  can  be termed as the modified equation of state of this fluid
	where as before  $ E_i=\rho _i v $ and $E_{i 0}=\rho _{i 0}v_0$ .
	\subsection{ Determination of thermodynamic derivatives :~$C_p,C_v,K_S,K_T$ }
	Now, the motivation of the present work  is to find the criteria of thermodynamic stability of the  universe . So  the conditions which may be satisfied by the parameters , indicate the thermodynamic features of the  evolution of the expanding  universe. For any system the thermodynamic derivatives i.e. the heat capacities,compressibility and isobaric expansibility determine whether the system is  stable  or not. So it  will be very important to analyze these parameters for each fluid in the present context.\par From the 1st law of thermodynamics in equation \,(\ref{8}), one can determine the two  heat capacities : heat capacity at constant volume $C_v$ and heat capacity at constant pressure $C_p$ by the relations ,
	\begin{eqnarray}
	C_v=\left [ \frac{\partial Q}{\partial T}\right ]_v =\left [\frac{\partial E}{\partial T}\right ]_v           \label{15} \\
	C_p=\left [\frac{\partial Q}{\partial T}\right ]_p=\left [\frac{\partial H}{\partial T}\right ]_p ,          \label{16}
	\end{eqnarray}
	where $H=E+pv$, is called the enthalpy of the system.
	\par Now the relation between temperature of a system at a constant volume $(T_a)$ and the physical equilibrium temperature is given by (according to \,\cite{Haba:2016bpt})
	\begin{equation}
	T=T_a \left (1+\frac{\gamma _i}{\rho _{i 0}}A_i \right )             \label{17}
	\end{equation}
	The temperature at a constant volume $T_a$ also evolves as
	\begin{equation}
	T_a=T_0 a^{-3\omega _i} .                     \label{18}
	\end{equation} So from  equations \,(\ref{17}) and  \,(\ref{18}),
	\begin{equation}
	E_i=E_{i 0} \left (\frac{1}{T_0}\right )T^{\frac{2+\omega _i}{1+\omega _i}}T_a^{-\frac{1}{1+\omega _i}}.              \label{19}
	\end{equation}
	Now from equations \,(\ref{15})and  \,(\ref{19}) one obtains,
	\begin{equation}
	C_{i v}=\frac{2+\omega _i}{1+\omega _i}\left (\frac{E_{i 0}}{T_0}\right ){\left (\frac{T}{T_a} \right )}^{\frac{1}{1+\omega _i}},           \label{20}
	\end{equation} 
	and  equations \,(\ref{1}) ,(\ref{16}) and \,(\ref{19}) yield ( using the definition of enthalpy) 
	\begin{equation}
	C_{i p}=(2+\omega _i)\left (\frac{E_{i 0}}{T_0}\right ){\left (\frac{T}{T_a}\right )}^{\frac{1}{1+\omega _i}}\left [1-\frac{1}{2+\omega _i} \left  (\frac{T}{T_a}\right )\left (\frac{\partial {T_a}}{\partial T} \right )_{p_i}\right ].       \label{21}
	\end{equation}
	Also equation \,(\ref{21}) can be rewritten (using \ref{17}) as
	\begin{equation}
	C_{i p}=\frac{E_0}{T_0}\left (\frac{T}{T_a}\right )^{\frac{1}{1+\omega _i}}\left [(1+\omega _i)+T \left \{ \frac{\partial}{\partial T}\ln \left  (1+\frac{\gamma _i}{\rho _{i 0}}A_i \right ) \right  \} _{p_i} \right ] .       \label{22}
	\end{equation}
	Further assuming temperature and pressure as independent thermodynamic variables , the variation of volume can be expressed as (See ref. \cite{Barboza:2015rsa} and \cite{Bhandari:2017cow})
	\begin{equation}
	dv=v(\alpha dT-K_T dp)     \label{23}
	\end{equation}where
	\begin{equation}
	\alpha = \frac{1}{v}\left(\frac{\partial v}{\partial T}\right)_p       \label{24} 
	\end{equation}
	is familiar as thermal expansibility. The isothermal compressibility $(K_T)$ is given by
	\begin{equation}
	K_T=-\frac{1}{v}\left (\frac{\partial v}{\partial p}\right )_T             \label{25}.
	\end{equation}
	Similarly the adiabatic compressibility is 
	\begin{equation}
	K_S = -\frac{1}{v} \left (\frac{\partial v}{\partial p}\right  )_S   .       \label{26}
	\end{equation}
	In isothermal process,
	\begin{equation}
	\frac{\alpha}{K_T}=\left (\frac{\partial p}{\partial T}\right )_v   .      \label{27}
	\end{equation}
	Also one  has  the well established relation amongst heat capacities and compressibilities as \cite{Barboza:2015rsa}
	\begin{equation}
	\frac{C_p}{C_v}=\frac{K_S}{K_T} .      \label{28}
	\end{equation}   
	So one  can find the expression for isothermal expansibility $\alpha$ from equation \,(\ref{11}) as, 
	\begin{equation}
	\alpha _i =-\frac{1}{\omega _i T}+\frac{1}{1+\omega _i } \left [\frac{\partial}{\partial T} \left \{ \ln \left ( 1+\frac{\gamma _i}{\rho _{i 0}} A_i\right) \right \} \right]_{p_i}.
	\label{29}\end{equation}
	Now equations \,(\ref{27}) and \,(\ref{28}) yield 
	\begin{eqnarray}
	K_{i T}=\frac{\alpha _i v}{\omega _i C_{i v}}        \label{30} \\ 
	K_{i S}=\frac{\alpha _i v}{\omega _i C_{i p}}      .  \label{31}
	\end{eqnarray}
	Hence one obtains  the expressions for compressibilities from equations \,(\ref{29}) , (\ref{30}) and (\ref{31}) containing  the dependence of diffusion parameter $\gamma _i$ as,
	\begin{eqnarray}
	K_{i T}=\frac{v}{\omega _i C_{i v}}\left [-\frac{1}{\omega _i T} +\frac{1}{1+\omega _i}\left \{ \frac{\partial}{\partial T} \ln \left (1+\frac{\gamma _i}{\rho _{i 0}}A_i      \right )       \right \}_{p_i}   \right ]        \label{32}   \\
	K_{i S}=\frac{v}{\omega _i C_{i p}}\left [-\frac{1}{\omega _i T} +\frac{1}{1+\omega _i}\left \{ \frac{\partial}{\partial T} \ln \left (1+\frac{\gamma _i}{\rho _{i 0}}A_i      \right )       \right \}_{p_i}   \right ].        \label{33}
	\end{eqnarray}
	\subsection{Stability Conditions for cosmic  fluids  :}
	For thermodynamic stability of any fluid , it must follow the conditions \cite{Barboza:2015rsa,Bhandari:2017cow}  namely 
	$C_p ,C_v ,K_ T, K_S \geq 0$ . For the present work, using the expressions of these thermodynamical parameters , the stability conditions are presented in the following table \ref{tab:1} :
	\begin{center}
		\tiny
		\begin{table}[ht]
			\renewcommand{\arraystretch}{1.5}
			\caption{Conditions for Stability} \label{tab:1}
			\begin{tabular}{| >{\centering\arraybackslash}m{3cm}|>{\centering\arraybackslash}m{15cm}|}
				\hline
				{\bf range of } $\omega _i $  & {\bf Stability Condition}\\
				\hline
				$\omega_i \geq0$ &$ \left [\frac{\partial \left \{\ln \left (1+\frac{\gamma _i}{\rho _{i 0}}A _i    \right) \right \} }{\partial \ln T}      \right]_{p_i}  \geq \frac{1+\omega _i}{\omega _i } $\\
				\hline
				$-1\leq \omega_i \leq0$  & $ -(1+\omega _i) \leq \left [\frac{\partial \left \{\ln \left (1+\frac{\gamma _i}{\rho _{i 0}}A _i    \right) \right \} }{\partial \ln T}      \right]_{p_i} $\par $\mbox{and}  $\par $ ~~~~~~~~~~~~~~~~~~~~~\left [\frac{\partial \left \{\ln \left (1+\frac{\gamma _i}{\rho _{i 0}}A _i    \right) \right \} }{\partial \ln T}      \right]_{p_i}  \leq \frac{1+\omega _i}{\omega _i } ~~\mbox{simultaneously,}~~~~  $\par $\mbox{which is not possible hence unstable.} $ \\
				\hline
				\hspace{5mm}$-2 \leq \omega_i \leq -1$   &  Unstable . \\
				\hline
				\hspace{5mm}$\omega _i \leq -2 $ & $\left [\frac{\partial \left \{\ln \left (1+\frac{\gamma _i}{\rho _{i 0}}A _i    \right) \right \} }{\partial \ln T}      \right]_{p_i}  \geq -(1+\omega _i) $ \\
				\hline
				
			\end{tabular}
		\end{table}
	\end{center}
	
	\begin{figure}[H]
		\begin{center}
			\includegraphics[width=0.6\textwidth]{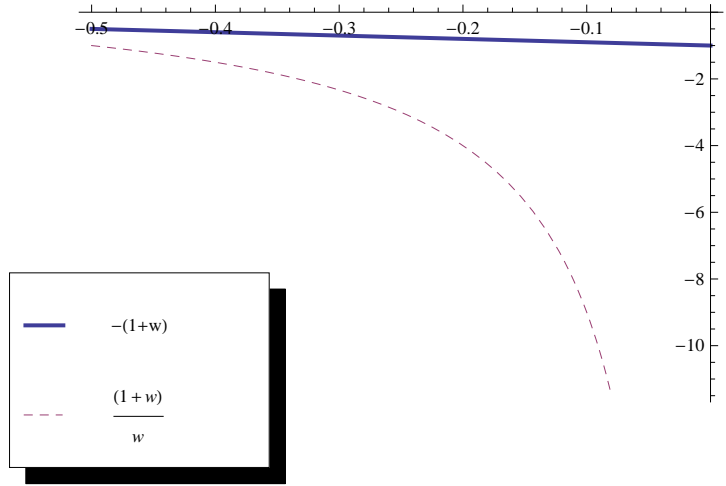}
			\caption{No point having value greater than $-(1+\omega_i)$ and less than $\frac{1+\omega_i}{\omega_i}$ simultaneously.}
			\label{fig:1}
		\end{center}
	\end{figure}

	\section{Stability criteria with variable equation of state parameter}     \label{sec3}
	So far we have examined the conditions for stability of cosmic fluids with constant equation of state parameters. In the present section the stability conditions will be discussed for cosmic fluids having  variable  equation of state parameters . The fluids are diffusive as before with no interaction amongst them. In such context, the solution of the conservation equation \,(\ref{6}) yields,
	\begin{equation}
	\rho_i=\left[\rho_{i0} + \gamma_i \int_{t_0}^t a^{-3} e^{F_i(a)}dt \right] e^{-F_i(a)}             \label{34}
	\end{equation}
	where, \begin{equation}
	F_i(a)=\int (1+\omega_i)~d(ln v)           \label{35}.
	\end{equation}
	Taking $$\int_{t_0}^t a^{-3} e^{F_i(a)}dt=A_i (t),$$ one can write equation \,(\ref{34}) as 
	\begin{equation}
	\rho_i= e^{-F_i(a)} \left[\rho_{i0} + \gamma_i A_i (t) \right]               \label{36}.
	\end{equation}\\
	\subsection{Derivation of the relation between physical equilibrium temperature (T) and temperature at constant volume $(T_a)$:}
	From 1st law of thermodynamics (i.e. equation \,(\ref{8})), by choosing  entropy (S) as a function of two independent variables volume (v) and temperature (T), the condition for dS to be  an exact differential gives 
	\begin{eqnarray}
	\left (\frac{\partial S}{\partial v}\right ) _ T &=&\frac{\rho}{T} (1+\omega)      \label{37}    \\
	\left (\frac{\partial S}{\partial T} \right ) _v &=&\frac{v}{T} \frac{d \rho}{d T}      \label{38}        \\ \mbox{and} ~~
	\frac{d \rho}{\rho} &=&\frac{1+ \omega }{\omega} \frac{d T}{T} - \frac{d \omega}{\omega} .        \label{39}
	\end{eqnarray}

	Now, from equation \,(\ref{39}) one obtains the  energy density as a function of  temperature as 
	\begin{equation}
	\rho=\frac{B}{\omega} e^{\int \frac{1+\omega}{\omega} \frac{dT}{T}}      \label{40}.
	\end{equation}
	Again, combining  equations \,(\ref{38}) and \, (\ref{40})  the expression for  entropy can be written as, 
	\begin{equation}
	S=v~\frac{1+\omega}{\omega T_a} B e^{\int \frac{1+\omega}{\omega} \frac{dT_a}{T_a}} .      \label{41}
	\end{equation}
	Now, as  for an adiabatic system  $\Delta S=0$, so one  can determine from equation \,(\ref{41}), the relation between $T_0$ and $T_a$ (i.e. evolution of $T_a$) as 
	\begin{equation}
	T_a  e^{-{\int_{present}^{t} \frac{1+\omega}{\omega} \frac{dT_a}{T_a}}}= T_0 \frac{1+\omega}{\omega} \frac{\omega_0}{1+\omega_0} a^3  .    \label{42}
	\end{equation}
	Now combining equations \,(\ref{35}) , \,(\ref{36}) and \,(\ref{42}) one obtains ,
	\begin{equation}
	\rho _i =\frac{1}{\omega _i} e^ {\int_{present}^{t}\frac{1+\omega _i}{\omega _i} \frac{d T_a}{T_a}} [\rho _{i 0} +\gamma _i A_i (t)]  ,      \label{43}
	\end{equation}
or equivalently 
	\begin{equation}
	\rho _i =\frac{1}{\omega _i} e^ {\int_{present}^{t}\frac{1+\omega _i}{\omega _i} \frac{d T_a}{T_a}} 3z_iB(A_i){v_0}^{- 1}        \label{44}
	\end{equation}
	$\mbox{~~where~~} z_i=\frac{\rho _{i 0} v_0}{3T_0} \mbox{~~and~~} B(A_i)=T_0+\frac{\gamma _i T_0}{\rho _{i 0}}A_i (t)$.
	
	Now as the adiabatic contribution of entropy for the i-th fluid (excluding the diffusive entropy $(S_{i d}=\int\frac{v_o \gamma _i}{T} d T))$ namely,
	\begin{equation}
	S_{i (ad)}=\frac{(1+\omega _i) \rho_i v}{T}      \label{45}
	\end{equation}
	 does not vary (i.e. $d \ln (S_{i (ad)}) = 0$) , so temperature can be written as,
	\begin{equation}
	T=\frac{1+\omega _{i 0}}{\omega _{i 0}L} \left (\frac{T_a}{T_0} \right) B(A_i)        \label{46}
	\end{equation}
	where  , $L$ is a constant . Now combining \,(\ref{45}) and \,(\ref{46}) yields
	\begin{equation}
	\left (1+\frac{\gamma _i}{\rho _{i 0}} A_i \right ) =\left (\frac{T}{T_a}\right ) Q     \label{47}
	\end{equation}
	with $Q=\frac{\omega _{i 0} L}{1+\omega _{i 0}}$, a constant .
	\subsection{Thermodynamic derivatives }
	  Similar to the above section , one can determine  the heat capacities and compressibilities  of the present thermodynamical system as follows: now  the solution of equation  \;(\ref{6}),(or equivalently equation \,( \ref{34}))  can be expressed as
	\begin{equation}
	d \ln \rho _i =-(1+ \omega _i) d\ln v+d\ln \left [1+\frac{\gamma _i}{\rho {i 0}}A_i \right ]   \label{48}.
	\end{equation}
	Again from equation \,(\ref{39}), one  finds 
	\begin{equation}
	d\ln \rho_i =\frac{1+\omega _i}{\omega _i} d\ln T -\frac{d \omega}{\omega}        \label{49}
	\end{equation}
	Combining the above equations \,(\ref{48}) and \,(\ref{49}) one can write, 
	\begin{equation}
	\frac{T}{E_i(1+\omega _i)}=Ce^{-\int \frac{1}{1+\omega _i } d\ln \left [1+\frac{\gamma _i}{\rho _{i 0}} A_i\right ]}    .   \label{50}
	\end{equation} 
	Now imposing the initial conditions , one  can formulate the modified equation of state in terms of evolution of energy with time and temperature  as 
	\begin{equation}
	E_i=E_{i 0}\left (\frac{1+\omega _{i 0}}{1+\omega _i}\right ) \left (\frac{T}{T_0}\right )e^{\int_{present}^{t}\frac{1}{1+\omega _i}d \ln \left (1+\frac{\gamma _i}{\rho _{i 0}}A_i\right )}      \label{51}.
	\end{equation} 
	The above  equation \,(\ref{51}) can be written according to equation \,(\ref{47}) as
	\begin{equation}
	E_i=E_{i 0}\left (\frac{1+\omega _{i 0}}{1+\omega _i}\right ) \left (\frac{T}{T_0}\right )e^{\int_{present}^{t}\frac{1}{1+\omega _i}d \ln \left (\frac{T}{T_a}\right )}       \label{52}.     
	\end{equation}
	So  using equations \,(\ref{15}) , \,(\ref{16}), \,(\ref{47}) and\,(\ref{51}) ,one  can easily express the heat capacities as 
	\begin{eqnarray}
	C_{i v}&=&\frac{E_i}{T(1+\omega _i)}\left [(2+\omega _i)-T\frac{d \omega}{d T}\right ]     \label{53}  \\
	C_{i p}&=&\frac{E_i}{T}\left [(1+\omega _i)+T \left \{\frac{\partial}{\partial T}\ln \left(1+\frac{\gamma _i}{\rho _{i 0}}A_i\right) \right  \}_p \right ]      \label{54}
	\end{eqnarray}
and hence one obtains the relation
	\begin{equation}
	\left (\frac{\partial p}{\partial T}\right )_v=\frac{\omega}{v} C_v +\rho \frac{d \omega}{d T}.     \label{55}
	\end{equation}
	Again equations \,(\ref{48}) and \,(\ref{49}) yield 
	\begin{equation}
	d\ln T =\frac{d \omega _i}{1+ \omega _i} -\omega _i d\ln v +\frac{\omega _i}{1+\omega _i} d \ln \left (1+\frac{\gamma _i}{\rho _{i 0}} A_i \right )    \label{56},
	\end{equation}
	which implies  along with equation \,(\ref{47}) that 
	\begin{equation}
	\left(\frac{\partial \omega}{\partial T}\right)_v =\frac{1+\omega _i + \omega _i ^2}{T(1+\omega _i)}  .    \label{57}
	\end{equation}
	So  from equations \,(\ref{27}) , \,(\ref{53}) , \,(\ref{55})  and \,(\ref{57}) one writes  
	\begin{equation}
	K_{i T}=\frac{\alpha _i v[2+\omega _i -T\frac{d \omega_i}{d T}]}{ C_{i v}[\omega _i \{ 1+(2+\omega _i -T\frac{d \omega _i }{d T})          \}  +1 +\omega _i ^2                                   ]                    }        \label{58}
	\end{equation}
	and  from (\ref{28}),one obtains  
	\begin{equation}
	K_{i S}=\frac{\alpha _i v[2+\omega _i -T\frac{d \omega_i}{d T}]}{ C_{i p}[\omega _i \{ 1+(2+\omega _i -T\frac{d \omega _i }{d T})          \}  +1 +\omega _i ^2                                   ]                    }       . \label{59}
	\end{equation}
	Now, using equation  (\ref{56})  the expression for $\alpha_i$ takes the form  
	\begin{equation}
	\alpha_i= \left [\frac{\partial(lnV)}{\partial T} \right ]_{p_i} = \frac{1}{\omega _i (1+\omega _i)} \left [\omega _i  \left \{\frac{\partial }{\partial T} \ln \left (1+\frac{\gamma _i}{\rho _{i 0}}A_i\right ) \right  \} _{p_i}+\frac{d\omega_i}{dT} -\frac{1+\omega_i}{T} \right ]       \label{60}
	\end{equation}
	and  the expressions for $K_T$ and $K_S$ are given by  (using equations \,(\ref{58}) , \,(\ref{59}) and  \,(\ref{60}))
	
	\begin{eqnarray}
	K_{iT}&=&\frac{V\left[\omega_i\{\frac{\partial}{\partial_T}ln \left(1+\frac{\gamma_i}{\rho_{i0}}A_i\right)\}_{p_i}+\frac{d\omega_i}{dT}-\frac{1+\omega_i}{T}   \right][2+\omega _i -T\frac{d \omega_i}{d T}]}{\omega_i (1+\omega_i) C_{iv}[\omega _i \{ 1+(2+\omega _i -T\frac{d \omega _i }{d T})          \}  +1 +\omega _i ^2]  }                  \label{61}   \\
	\mbox{and} ~~
	K_{iS}&=&\frac{V\left[\omega_i\{\frac{\partial}{\partial_T}ln \left(1+\frac{\gamma_i}{\rho_{i0}}A_i\right)\}_{p_i}+\frac{d\omega_i}{dT}-\frac{1+\omega_i}{T}   \right][2+\omega _i -T\frac{d \omega_i}{d T}]}{\omega_i (1+\omega_i) C_{ip}[\omega _i \{ 1+(2+\omega _i -T\frac{d \omega _i }{d T})          \}  +1 +\omega _i ^2]  }          .     \label{62}
	\end{eqnarray} 
	Now according to the conditions of thermodynamic stability , we analyze the equations \,(\ref{53}) , \,(\ref{54}) , \,(\ref{61}) and \,(\ref{62}) to find the restrictions of thermodynamic stability in different ranges of $\omega _i$ in the following  Table \ref{tab:2} .
	
	\begin{center}
		\tiny
		\begin{table}[ht]
			\renewcommand{\arraystretch}{1.5}
			\caption{Conditions for Stability Criteria} \label{tab:2}
			\begin{tabular}{| >{\centering\arraybackslash}m{2cm}|>{\centering\arraybackslash}m{16cm}|}
				\hline
				{\bf range of } $\omega _i $  & {\bf Stability Condition}\\
				\hline
				$\omega_i \geq0$ &$ \frac{d\ln \omega_i}{d \ln T}\leq \frac{2+\omega_i}{\omega_i}~~ 
				\mbox{and} \left [\frac{\partial \left \{\ln \left (1+\frac{\gamma _i}{\rho _{i 0}}A _i    \right) \right \} }{\partial \ln T}      \right]_{p_i} \geq max \left[-(1+\omega _i) ,\left ( \frac{1+\omega_i }{\omega_i}-\frac{d \ln \omega_i}{d \ln T}\right )\right]$\\
				\hline
				$-1\leq \omega_i \leq0$  & ~~~~~ $\frac{d\ln \omega_i}{d \ln T}\geq \frac{2+\omega_i}{\omega_i} ~~\mbox{and}$\par$~~\left [\frac{\partial \left \{\ln \left (1+\frac{\gamma _i}{\rho _{i 0}}A _i    \right) \right \} }{\partial \ln T}      \right]_{p_i} \geq max \left[-(1+\omega _i) ,\left ( \frac{1+\omega_i }{\omega_i}-\frac{d \ln \omega_i}{d \ln T}\right )\right]~ \mbox{if}~\frac{d \ln \omega_i}{d \ln T} \leq \frac{1+3\omega_i +2\omega_i^2}{\omega_i^2}. $\par$\mbox{Again}-(1+\omega _i) \leq  \left [\frac{\partial \left \{\ln \left (1+\frac{\gamma _i}{\rho _{i 0}}A _i    \right) \right \} }{\partial \ln T}      \right]_{p_i} \leq \left ( \frac{1+\omega_i }{\omega_i}-\frac{d \ln \omega_i}{d \ln T}\right )~~ \mbox{if}~~ \frac{d \ln \omega_i}{d \ln T} \geq \frac{1+3\omega_i +2\omega_i^2}{\omega_i^2}~~~~~~~~~  $ \\

				\hline
				\hspace{5mm}$\omega _i \leq -1 $ & $ \frac{d\ln \omega_i}{d \ln T}\leq \frac{2+\omega_i}{\omega_i} ~~~~\mbox{and} ~~~\left [\frac{\partial \left \{\ln \left (1+\frac{\gamma _i}{\rho _{i 0}}A _i    \right) \right \} }{\partial \ln T}      \right]_{p_i} \geq max \left[-(1+\omega _i) ,\left ( \frac{1+\omega_i }{\omega_i}-\frac{d \ln \omega_i}{d \ln T}\right )\right] $ \\
				\hline
				
			\end{tabular}
		\end{table}
	\end{center}

	 \begin{figure}[H]
	 	\begin{center}
	 		\includegraphics[width=1.2\textwidth]{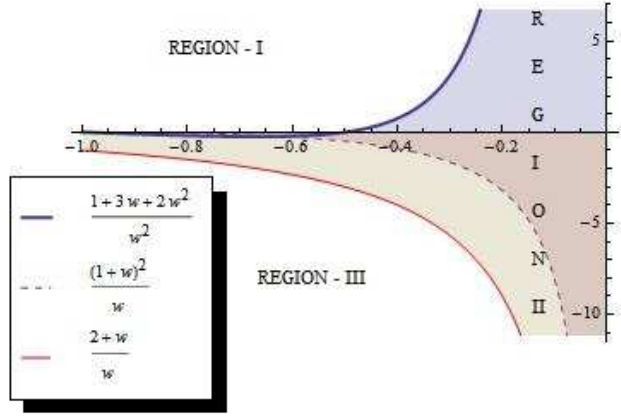}
	 		\caption{For the values of $\frac{d \ln \omega_i}{d \ln T} ~~ \epsilon$ REGION-I, the stability condition is $-(1+\omega _i) \leq  \left [\frac{\partial \left \{\ln \left (1+\frac{\gamma _i}{\rho _{i 0}}A _i    \right) \right \} }{\partial \ln T}      \right]_{p_i} \leq \left ( \frac{1+\omega_i }{\omega_i}-\frac{d \ln \omega_i}{d \ln T}\right )$,~~for~$\frac{d \ln \omega_i}{d \ln T} ~ \epsilon$ REGION-II, the stability condition is $\left [\frac{\partial \left \{\ln \left (1+\frac{\gamma _i}{\rho _{i 0}}A _i    \right) \right \} }{\partial \ln T}      \right]_{p_i} \geq max \left[-(1+\omega _i) ,\left ( \frac{1+\omega_i }{\omega_i}-\frac{d \ln \omega_i}{d \ln T}\right )\right]$ and in REGION-III unstable. }
	 		
	 		\label{fig:2}
	 	\end{center}
	 \end{figure}

 \begin{center}
 	\begin{table}[ht]
 		\renewcommand{\arraystretch}{1.5}
 		\caption{Conclusion on the result in Dark energy range ($-0.33<\omega_i<0 $)} \label{tab:3}
 		\begin{tabular}{| >{\centering\arraybackslash}m{2cm}|>{\centering\arraybackslash}m{10cm}|>{\centering\arraybackslash}m{5cm}|}
 			\hline
 			{\bf Nature of Dark Energy} & {\bf Constraint on diffusion parameter $\gamma_i$}  & {\bf Restriction on $\frac{d\ln \omega_i}{d \ln T} $ from graphs}\\
 			\hline
 			Constant equation of state parameter & Always unstable & Always unstable  \\
 			\hline
 		Variable equation of state parameter & $\left [\frac{\partial \left \{\ln \left (1+\frac{\gamma _i}{\rho _{i 0}}A _i    \right) \right \} }{\partial \ln T}      \right]_{p_i} \geq max \left[-(1+\omega _i) ,\left ( \frac{1+\omega_i }{\omega_i}-\frac{d \ln \omega_i}{d \ln T}\right )\right]$ & $\frac{2+\omega_i}{\omega_i} \leq \frac{d \ln \omega_i}{d \ln T} \leq \frac{1+3\omega_i+2\omega_i^2}{\omega_i^2}$\\
 			\hline
 			
 		\end{tabular}
 	\end{table}
 \end{center}

\section{brief discussions}     \label{sec4}
A detailed thermodynamic study of non-interacting cosmic fluids  having diffusive nature has been organized in the present work. As a result,  individual fluid  does not obey the energy conservation relation ,rather the non-conservation is proportional to the corresponding diffusive current. Each component of the cosmic fluids is assumed to have barotropic equation of state with constant or variable equation of state parameter. The stability conditions are  expressed in the form of inequalities for different ranges of equation of state parameter.  Also, the  stability conditions for both constant (Table \ref{tab:1} )and variable (Table \ref{tab:2}) equation of state parameter in the range $-1<\omega_i <0$ (i.e. for dark energy region) have been presented graphically in figure \ref{fig:1} and \ref{fig:2} respectively. From figure \ref{fig:2}, different restrictions on diffusion parameter $\gamma_i$  are found depending on whether the values of $\frac{d \ln \omega_i}{d \ln T}$ (i.e. relative variation of $\omega_i$ with respect to the temperature T) belong to REGION-I or REGION-II . However due to expanding nature of the universe, the isobaric expansibility ($\alpha_i$) should always be positive. This condition restricts the function $\frac{d \ln \omega_i}{d \ln T}$ only in REGION-II,which is acceptable for expanding universe (Table \ref{tab:3}).
	
	\par Further, it is interesting to note that, for constant equation of state parameter, (i.e. Table \ref{tab:1} ), the system cannot be thermodynamically stable for $-1<\omega_i\leq0$.
	 This result is very much similar to that of \cite{Barboza:2015rsa}. However, the present work is thermodynamically stable in phantom region  in contrary with non-diffusive cosmic fluids \cite{Barboza:2015rsa}. Finally, it is found that diffusive fluid with variable equation of state parameter is thermodynamically stable for all possible values of $\omega_i$ under some restrictions on the diffusion parameter $\gamma_i$ and nature of variation of $\omega_i$ with respect to temperature.

	\section{Acknowledgement}
	 
	 The authors are thankful to IUCAA, Pune, India for research facilities at Library. The author SM acknowledges UGC-JRF and  PB acknowledges DST-INSPIRE  (File no: IF160086) for awarding Research fellowship. Also  SC acknowledges the UGC-DRS Program in the Department of Mathematics, Jadavpur University.\\\\

	.
	
\end{document}